\documentstyle[twoside,fleqn,espcrc2]{article}
\newread\epsffilein    % file to \read
\newif\ifepsffileok    % continue looking for the bounding box?
\newif\ifepsfbbfound   % success?
\newif\ifepsfverbose   % report what you're making?
\newdimen\epsfxsize    % horizontal size after scaling
\newdimen\epsfysize    % vertical size after scaling
\newdimen\epsftsize    % horizontal size before scaling
\newdimen\epsfrsize    % vertical size before scaling
\newdimen\epsftmp      % register for arithmetic manipulation
\newdimen\pspoints     % conversion factor
\def\epsfbox#1{\global\def\epsfllx{72}\global\def\epsflly{72}%
   \global\def\epsfurx{540}\global\def\epsfury{720}%
   \def\lbracket{[}\def\testit{#1}\ifx\testit\lbracket
   \let\next=\epsfgetlitbb\else\let\next=\epsfnormal\fi\next{#1}}%
\def\epsfgetlitbb#1#2 #3 #4 #5]#6{\epsfgrab #2 #3 #4 #5 .\\%
   \epsfsetgraph{#6}}%
\def\epsfnormal#1{\epsfgetbb{#1}\epsfsetgraph{#1}}%
\def\epsfgetbb#1{%
\openin\epsffilein=#1
\ifeof\epsffilein\errmessage{I couldn't open #1, will ignore it}\else
   {\epsffileoktrue \chardef\other=12
    \def\do##1{\catcode`##1=\other}\dospecials \catcode`\ =10
    \loop
       \read\epsffilein to \epsffileline
       \ifeof\epsffilein\epsffileokfalse\else
          \expandafter\epsfaux\epsffileline:. \\%
       \fi
   \ifepsffileok\repeat
   \ifepsfbbfound\else
    \ifepsfverbose\message{No bounding box comment in #1; using defaults}\fi\fi
   }\closein\epsffilein\fi}%
\def\epsfclipstring{}% do we clip or not?  If so,
\def\epsfsetgraph#1{%
   \epsfrsize=\epsfury\pspoints
   \advance\epsfrsize by-\epsflly\pspoints
   \epsftsize=\epsfurx\pspoints
   \advance\epsftsize by-\epsfllx\pspoints
   \epsfxsize\epsfsize\epsftsize\epsfrsize
   \ifnum\epsfxsize=0 \ifnum\epsfysize=0
      \epsfxsize=\epsftsize \epsfysize=\epsfrsize
      \epsfrsize=0pt
     \else\epsftmp=\epsftsize \divide\epsftmp\epsfrsize
       \epsfxsize=\epsfysize \multiply\epsfxsize\epsftmp
       \multiply\epsftmp\epsfrsize \advance\epsftsize-\epsftmp
       \epsftmp=\epsfysize
       \loop \advance\epsftsize\epsftsize \divide\epsftmp 2
       \ifnum\epsftmp>0
          \ifnum\epsftsize<\epsfrsize\else
             \advance\epsftsize-\epsfrsize \advance\epsfxsize\epsftmp \fi
       \repeat
       \epsfrsize=0pt
     \fi
   \else \ifnum\epsfysize=0
     \epsftmp=\epsfrsize \divide\epsftmp\epsftsize
     \epsfysize=\epsfxsize \multiply\epsfysize\epsftmp   
     \multiply\epsftmp\epsftsize \advance\epsfrsize-\epsftmp
     \epsftmp=\epsfxsize
     \loop \advance\epsfrsize\epsfrsize \divide\epsftmp 2
     \ifnum\epsftmp>0
        \ifnum\epsfrsize<\epsftsize\else
           \advance\epsfrsize-\epsftsize \advance\epsfysize\epsftmp \fi
     \repeat
     \epsfrsize=0pt
    \else
     \epsfrsize=\epsfysize
    \fi
   \fi
   \ifepsfverbose\message{#1: width=\the\epsfxsize, height=\the\epsfysize}\fi
   \epsftmp=10\epsfxsize \divide\epsftmp\pspoints
   \vbox to\epsfysize{\vfil\hbox to\epsfxsize{%
      \ifnum\epsfrsize=0\relax
        \includegraphics{#1}%
      \else
        \epsfrsize=10\epsfysize \divide\epsfrsize\pspoints
        \includegraphics{#1}%
      \fi
      \hfil}}%
\global\epsfxsize=0pt\global\epsfysize=0pt}%
{\catcode`\%=12 \global\let\epsfpercent=%\global\def\epsfbblit{%BoundingBox}}%
\long\def\epsfaux#1#2:#3\\{\ifx#1\epsfpercent
   \def\testit{#2}\ifx\testit\epsfbblit
      \epsfgrab #3 . . . \\%
      \epsffileokfalse
      \global\epsfbbfoundtrue
   \fi\else\ifx#1\par\else\epsffileokfalse\fi\fi}%
\def\epsfempty{}%
\def\epsfgrab #1 #2 #3 #4 #5\\{%
\global\def\epsfllx{#1}\ifx\epsfllx\epsfempty
      \epsfgrab #2 #3 #4 #5 .\\\else
   \global\def\epsflly{#2}%
   \global\def\epsfurx{#3}\global\def\epsfury{#4}\fi}%
\def\epsfsize#1#2{\epsfxsize}
\let\epsffile=\epsfbox

\begin{document}

\title{Yang-Mills classical solutions and fermionic zero modes
        from lattice calculations \thanks{Presented by A. Montero at Lattice '97}}

\author{M. Garc\'{\i}a  P\'erez \address{ Institut f. Theoretische Physik,
                                         University of Heidelberg,
                                         D-69120 Heidelberg, Germany. },
        A. Gonz\'alez-Arroyo \address{ Departamento de F\'{\i}sica Te\'orica C-XI,
                                       Universidad Aut\'onoma de Madrid, Madrid 28049, Spain.} ,
        A. Montero${}^{\ \rm b}$
        and C. Pena${}^{\ \rm b}$}

\begin{abstract}

     We study a series of problems in classical 
Yang-Mills theories using lattice methods. We first investigate SU(N) self-dual configurations
on the torus with twisted boundary conditions. We also study the zero modes
of the Dirac equation in topological non-trivial background Yang-Mills fields.

\end{abstract}

\maketitle

\section{Introduction}

It  has long been argued that classical configurations might play a relevant role 
in the understanding of low-energy phenomena like confinement
and chiral symmetry breaking. Here we will present a study
of a particular kind of solutions which appear on the torus
when twisted boundary conditions are imposed. These are instanton-like configurations 
with the peculiarity of having fractional topological charge. In the large $N$ limit
their action scales as $8\pi^2/N g^2$. 
As has been argued in \cite{TA} they may be important to understand  
the long-distance dynamics of QCD. 
In this paper we will present a preliminary investigation of: i) the large $N$ 
limit behaviour  ii) the zero modes of
the Dirac operator on the background of fractional-charge instantons.

\section{Yang-Mills classical solutions}

The first part of this work is a study of $SU(N)$ (anti) self-dual Yang-Mills 
classical solutions on the torus with twisted boundary conditions. It is known
that in this case the topological charge Q is not always an integer \cite{Thoof},
but is given by 
\begin{equation}
 Q = \frac{1}{16 \pi^2} \int Tr \left( F_{\mu \nu} \tilde{F}_{\mu \nu} \right) d_4x = \nu - \frac{\kappa}{N}  ,
\end{equation}
\noindent
where $\nu$ and $\kappa$ are integers, $F_{\mu \nu}$ is the Yang-Mills strength 
tensor in the fundamental representation and $\tilde{F}_{\mu \nu}$ its dual. 
Here 
\begin{equation}
 \kappa = \frac{1}{4} n_{\mu \nu} \tilde{n}_{\mu \nu} = \vec{k} \hspace{0.02 in} \vec{m} ,
\end{equation}

\noindent
with $n_{\mu \nu}$ the usual twist tensor and $k_i=n_{0i}$, $n_{ij}=\epsilon_{ijk}m_k$.

From Schwarz-inequality
\begin{equation}
S = \frac{1}{2} \int Tr \left( F_{\mu \nu} F_{\mu \nu} \right) d_4x  \ge 8 \pi^2 |Q| 
\end{equation}

\noindent
one easily derives that 
if $\kappa \ne 0 \ ({\rm modulo} \  N)$ an obstruction is found for zero-action configurations.
We are interested in those solutions with minimal non-trivial action
\begin{equation}
S = 8 \pi^2 |Q| = \frac{8 \pi^2}{N}, 
\end{equation}
on a volume $[0,L]^3 \times [0,T]$, with $T \gg L$. 
Some of them are already known: i) For specific values of $N$, 't Hooft has
constructed non-abelian solutions with constant field strength which turn
out to be (anti)self-dual whenever the sides of the torus satisfy certain
relations (see \cite{Thoof1} for details), ii)
 there are also numerical studies
of the solution for $SU(2)$ with twist
$ \vec m = (1,1,1)$ and
$ \vec k = (1,1,1) $ \cite{Cool1}. The present work is inspired by this last reference.

We have restricted our analysis to the following  
twist tensors:
\begin{enumerate}
 \item Spatial twist, always $ \vec m = (1,1,1)$.
 \item Temporal twist, two cases:
  \begin{itemize}
   \item $ \vec k = (1,0,0) $ for $N=3,4,,...,13$ the solution is in this case antiself-dual, $Q=-1/N$.
   \item $ \vec k = (n,n,n) $ for $N=3n+1=4,7,$ $10,13$ the solution is here self-dual, $Q=1/N$.   
  \end{itemize}
\end{enumerate}

The solutions were generated on  $N_s^3\times N_t$ lattices ($N_t >> N_s$)
by cooling with the  Cabibbo-Marinari-Okawa algorithm \cite{cabibo}.
%\noindent
%the solutions have the following properties:
They verify  the following properties:
\begin{enumerate}
 \item Scaling towards the continuum solution.
 \item Self-duality or antiself-duality.
 \item The energy profile
  \begin{equation}
   \epsilon(t) = \int S(\vec x, t) d_3 \vec{x} 
  \end{equation}
   with $S(\vec x, t)$ the action density, is instantonic and independent of temporal twist.
 \item Behavior for large N:
  \begin{itemize}
   \item The energy profile behaves as $\epsilon(t) \simeq f(t/N)/N^2 $ as 
illustrated in Figure 1.
   \item The norm of the field strength tensor becomes independent of the spatial coordinates.   
   \item  In the gauge
\begin{equation}
 A_0 = 0 \hspace{.3 in}  A_i(t=-\infty,\vec x) = 0 ,
\end{equation}
one eigenvalue  (the same one for $A_i$, $B_i$, $E_i$) dominates the solution in the sense
that it gives the major contribution to the action.
  \end{itemize}
\end{enumerate}

A more detailed description of the properties of these solutions will be
presented in \cite{Cool3}.
\begin{figure}
\leavevmode
\epsfverbosetrue
\epsfxsize=210pt
\epsfysize=215pt
\epsffile{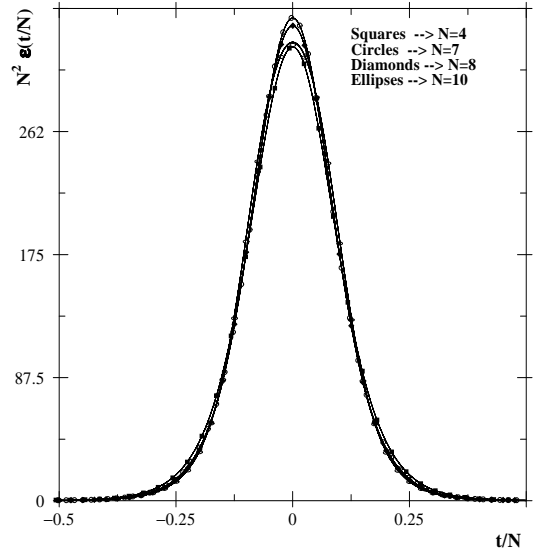}
\vspace{-1cm}
\caption{ The energy profile for 4 different $SU(N)$ solutions multiplied by $N^2$ is plotted
as a function of $t/N$. }
\end{figure}

\section{Fermionic zero modes}

The second part of this work is a study of fermionic zero modes of the Dirac equation
in the background of topologically non-trivial Yang-Mills fields.

To obtain the zero modes we have used the lattice approach with Wilson fermions and
also with naive (Kogut-Susskind) fermions.
In the massless case, the Dirac operator reads
\begin{eqnarray*}
\not\!\!D \Psi(n) = - \frac{1}{2} \sum_{\mu} [ (r-\gamma_{\mu}) U_{\mu}(n) \Psi (n+\hat{\mu}) +
\\           (r+\gamma_{\mu})  {U_{\mu}}^{\dag}(n-\hat{\mu}) \Psi (n-\hat{\mu}) ] + 4r\Psi(n),
\end{eqnarray*}
\noindent
where $\Psi$ is the fermion field, $U_{\mu}$ the gauge field,  $\gamma_{\mu}$ the
Dirac matrices and $r$ the Wilson parameter.

 We look for the lowest eigenvalues (and the corresponding eigenvectors) of the operator $(\gamma_5 \not\!\!D)^2 $ using three different methods:
\begin{itemize}
 \item Conjugate gradient
 \item Local minimization
 \item Lanczos (only for eigenvalues)
\end{itemize}
The results we have obtained are the same independently of the method employed in the
diagonalization.
We should point out that the Wilson parameter 
must be big enough to avoid fermionic doubling and, at the same time, small
enough such as not to distort the solution. 
For a lattice size $N_s^4$ with $N_s\geq 8$ a good value is $r=0.001$.

As a check of the method we successfully extracted the zero mode for the instanton
solution, also known as the 't Hooft zero mode. 

Here, we will present our results for the zero modes in the background of the SU(2) 
self-dual gauge field with $Q=1/2$ on the $[0,L]^4$ torus with twist  $\vec m = \vec k =(1,1,1)$.
Fermions are taken in the adjoint representation of $SU(2)$ (to avoid twist singularities).

\begin{table}
\caption{ The three lowest eigenvalues of $(\gamma_5 \not\!\!\!D)^2 $ in the presence of a $Q=1/2$ instanton on a
$N_s^4$ lattice.}
\begin{tabular}{||c|c|c|c||}
\hline
$N_s$  &  $ 1^{st} (\times 10^{-9})$ & $ 2^{nd} (\times 10^{-9})$  & $ 3^{rd} (\times 10^{-6})$ \\   
\hline \hline 
 11  &  6.946  &  6.946   &  4.150    \\ \hline
 12  &  4.835  &  4.835   &  4.140    \\ \hline
 13  &  3.546  &  3.546   &  4.107    \\ \hline 
 14  &  2.633  &  2.633   &  4.103    \\ \hline  
\end{tabular}
\end{table}

We have worked on $N_s^4$ lattices with $N_s=11,12,13,14$ and $r=0.001$. 
The three lowest eigenvalues obtained are shown in Table 1. We see that two
of these eigenvalues scale to zero. These would correspond to the two zero
modes predicted by the Index Theorem. They are related to each other by the operation
$\Psi_2 = \gamma_1 \gamma_3 \Psi_1^{ \star}$. 
Our results show that the gauge invariant density of these lowest eigenvectors
scales nicely towards a continuum curve. Furthermore, this density is cubic symmetric
and peaks at the position of the $Q=1/2$ instanton. This is examplified by
Figure 2, where we display the quantity:
\begin{equation}
\Phi (x_4) = \sum_{\alpha  i } \int {\Psi_{ \alpha i}}^* \Psi_{\alpha i} \hspace{0.1 in}dx_1 dx_2 dx_3 .
\end{equation}
In the above equation, $x_4$ is one of $\{x,y,z,t\}$, and $x_1,x_2,x_3$ are the other three components, where
the origin of coordinates is taken at the maximum of the gauge field configuration. It is clear from
the figure the nice scaling of the results. The curves for naive fermions are not displayed,
but fall on top of the other.

A more detailed description of these solutions and of others obtained
for different classical Yang-Mills backgrounds will be given in \cite{Modos}.

\begin{figure}
\leavevmode
\epsfverbosetrue
\epsfxsize=210pt
\epsfysize=215pt
\epsffile{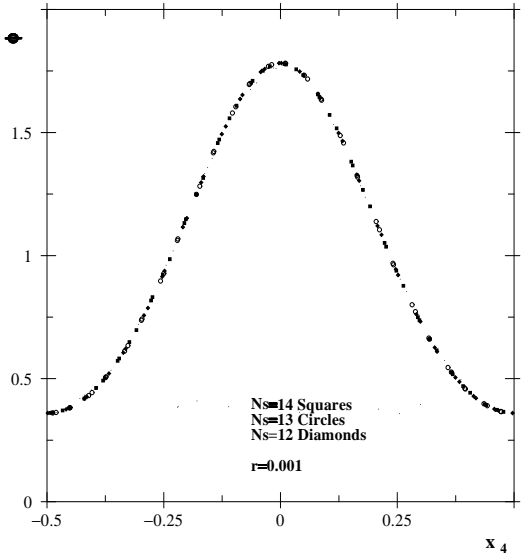}
\vspace{-1cm}
\caption{ The quantity $\Phi$ is plotted as a function of $x_4$ for different lattice sizes. }
\end{figure}

\vspace{5mm}
{\bf Acknowledgments}: DFG financial support for MGP is thankfully acknowledged. AGA and AM
would like to thank CICYT (Grant AEN $96-1664$) for their financial support.

\end{document}